\documentclass[a4paper,12pt]{iopart}
\usepackage{iopams,setstack}
\usepackage[utf8x]{inputenc}
\usepackage{epsf}%
\usepackage[pdftex]{graphicx}
\usepackage{amsfonts,latexsym}
\usepackage{amssymb,bm}
\usepackage{epsfig}
\usepackage{color}

\newcommand{\nuovertwo}{\case{\nu}{2}}
\newcommand{\half}{\case{1}{2}}
\newcommand{\AS}[1]{\cite[{\hspace{-1.5mm}\it #1}]{Abramowitz}}
\newcommand{\KummerU}[3]{U\left(#1,#2,#3\right)}
\newcommand{\KummerM}[3]{M\left(#1,#2,#3\right)}
\newcommand{\GAMMA}[1]{\Gamma\left(#1\right)}

\begin{document}

\title{Solution of the quantum harmonic oscillator plus a delta-function potential at the origin:
The {\it oddness} of its even-parity solutions}

\author{J.  Viana-Gomes and N. M. R. Peres  }
\address{University of Minho, Physics Department, CFUM, P-4710-057, Braga, Portugal}

\begin{abstract}
We derive the energy levels associated with the even-parity  wave functions 
of the harmonic oscillator with an
additional  
delta-function potential at the origin. Our results bring to the attention of students
a non-trivial and analytical example of a modification of the usual harmonic oscillator 
potential, with emphasis on the modification of the boundary conditions at the origin.
This problem  calls the attention 
of the students to an inaccurate statement
in quantum mechanics textbooks often found in the context of 
solution of the harmonic oscillator problem.
\end{abstract}


\section{Introduction}

Every single book on quantum mechanics gives the solution of the  harmonic oscillator problem.
The reasons for that are, at least, two: (i) it is a simple problem, amenable to
different methods of solution, such as the Frobenius method for solving
 differential equations \cite{Bell,Lebedev}, and  the algebraic method 
leading to the  introduction of  creation and annihilation operators \cite{Griffiths}.
This problem has therefore a natural pedagogical value; 
(ii)  the system itself
 has immense applications in different fields of physics and chemistry \cite{Bloch,Moshinsky} and it will
appear time and time again in the scientific life of a physicist.

Another problem often found in quantum mechanics textbooks is the calculation of the bound 
state (negative energy) of the potential $V(x)=\alpha\delta(x)$, with $\alpha<0$.
 The latter example is instructive for the
student because it cracks down the miss-conception that the continuity of
the wave function and its first derivative at an interface is the only possible boundary condition in quantum
problems. In fact, for this potential, the first derivative of the wave function is
discontinuous at $x=0$. To see the origin of this result, let us write the Schr\"odinger equation as
\begin{equation}
 -\frac{\hbar^2}{2m}\frac{d^2 \psi(x)}{dx^2}+\alpha\delta(x)\psi(x)=E\psi(x)\,.
\label{eq_Schrod_delta}
\end{equation}
Integrating Eq. (\ref{eq_Schrod_delta}) in an infinitesimal region around $x=0$ we get \cite{Griffiths}
\begin{equation}
 \lim_{\epsilon\rightarrow0}\left.\frac{d\psi(x)}{dx}\right|_{-\epsilon}^{+\epsilon}-\frac{2m\alpha}{\hbar^2}\psi(0)=0
      \Leftrightarrow
      \psi'_>(0)-\psi'_<(0)=\frac{2m\alpha}{\hbar^2}\psi(0),
\label{eq_BC_delta}
\end{equation}
where $\psi_\gtrless(x)$ are the wave functions in each side of the delta potential
  and we have used Newton's notation for derivatives where the primes over functions
 denote the order of the derivative of that function.
It is clear from Eq. (\ref{eq_BC_delta}) that the first derivative of the wave function
is discontinuous. An explicitly calculation gives the eigenstate of the system in the form 
\begin{equation}
 \psi(x)=\sqrt{\kappa}\,e^{-\kappa \vert x\vert },
 \label{psiU_m0}
\end{equation}
which has a kink at the origin  as well as a  characteristic length scale given by $\kappa=m\alpha/\hbar^2$. 
The eigenvalue associated with the wave function (\ref{psiU_m0}) 
is $E=-  \hbar^2\kappa^2/(2m)$.
When $E>0$, the system has only scattering states for both positive and negative values of $\alpha$.
We now superimpose  a harmonic potential potential on the already present delta-function potential. 
Due to the  confining harmonic potential, the new system only has
bound states, no mater the sign of $\alpha$ and $E$.
Thus, the problem we want to address is the calculation of the eigenstates and eigenvalues
of the potential
\begin{equation}
 V(x)=\frac{m}{2}\omega^2x^2+\alpha\delta(x)\,.
\end{equation}

As we will see, this problem has both  ``trivial'' and  non-trivial solutions. Furthermore, it allows 
a little excursion into the  world of special functions. Indeed, special functions play a prominent 
role in theoretical physics, to a point that the famous
 {\it Handbook of Mathematical Functions}, by Milton Abramowitz and Irene Stegun \cite{Abramowitze},
would be 
one of the three texts (together with the {\it Bible} and Shakespeare complete works)
 Michal Berry would take with him to a desert island \cite{Berry}.
In a time where symbolic computational software is becoming more and more the source 
of mathematical data, we hope 
with this problem to show that everything we need can be found the good old text 
of Milton Abramowitz and Irene Stegun \cite{Abramowitze}.

In addition, 
this problem will call the attention 
of  students to an inaccurate statement
in quantum mechanics textbooks often found in the context of the solution of the harmonic oscillator problem. Our approach is pedagogical, in the sense that illuminates the 
role of boundary conditions imposed on the even-parity wave functions by the $\delta-$function
potential.
(We note that after the submission of this article, the work
by  Busch {\it et al.} \cite{Busch1998}  was brought to our attention; see {\it note added}
at the end of article.)

\section{Solution of the harmonic oscillator with a delta-function}

The Hamiltonian of the system is 
\begin{equation}
 H= -\frac{\hbar^2}{2m}\frac{d^2}{dx^2}+\frac{m}{2}\omega^2x^2+\alpha\delta(x)\,.
\label{eq_Hamilt_HO}
\end{equation}
Following tradition, we introduce dimensionless variables using the intrinsic length scale of the problem
$a_0^2=\hbar/(m\omega)$. Making the substitution $y=x/a_0$ in Eq. (\ref{eq_Hamilt_HO}) 
we can write the Schr\"odinger equation as
\begin{equation}
\frac{d^2 \psi(y)}{dy^2}+(2\epsilon-y^2)\psi(y)+2g\delta(y)\psi(y)=0\,,
\label{eq_Hamilt_HO_dless}
\end{equation}
where $\epsilon=ma_0^2E/\hbar^2$ and $g=\alpha a_0m/\hbar^2$. With $g=0$, Eq. (\ref{eq_Hamilt_HO_dless})
is recognized as the Weber-Hermite differential equation \cite{Bell}. In quantum mechanics textbooks,
the solution of Eq. (\ref{eq_Hamilt_HO_dless}) with $g=0$ proceeds by making the substitution
 \begin{equation}
  \psi(y)=e^{-y^2/2}w(y)\,.
  \label{w_def}
 \end{equation}
At same time, it is a common practice to 
 write $2\epsilon=2\nu +1$, where $\nu$ is a real number.
This allow us to transform  Eq. (\ref{eq_Hamilt_HO_dless}) into
\begin{equation}
 w''-2yw'+2\nu w -2g\delta(y)w=0\,,
\label{eq_hermite} 
\end{equation}
which is Hermite's differential equation when $g=0$;
 a further substitution,  $z=y^2$, transforms the Eq.(\ref{eq_hermite}) into the Kummer's equation,  
\begin{equation}
    zw''+(b-z)w'-a\nu w=0~~~~~~\mathrm{with}~a=-\nuovertwo~\mathrm{and}~b=\half\,,
    \label{eq_kummer}
\end{equation}
which, obviously, has two linearly independent solutions: the confluent hypergeometric functions $M(a,b,z)$ and $U(a,b,z)$;
these functions 
also  known as  Kummer's functions (the latter solution is sometimes  referred as  Tricomi's function)
 \footnote{These functions may also be referred as the confluent hypergeometric
 function of the first and second kind, with the notation $M(a,b,z)=_1F_1(a;b;z)$ and $U(a,b,z)=z^{-a}_2F_0(a;1+a-b;-1/z)$.
  All these notations can be found when using computational methods and software.}.
Thus, the general solution of Eq. (\ref{eq_kummer}) is
\begin{equation}
    w(z)=A_\nu\,M(-\nuovertwo,\half,z)+B_\nu\,U(-\nuovertwo,\half,z),
    \label{sol_w}
\end{equation}
where $A_\nu$ and $B_\nu$ are arbitrary complex constants and $\nu$ is an arbitrary real number.
The $U(-\nuovertwo,\half,z)$ function can also be written in terms the functions $M(a,b,z)$ as~\AS{a}
  \begin{equation}
     \KummerU{\nuovertwo}{\half}{z}=\pi
                                     \left\{\frac{\KummerM{-\nuovertwo}{\half}{z}}{\GAMMA{\half}\GAMMA{\half-\nuovertwo}}-
                                     \sqrt{z}\frac{\KummerM{\half-\nuovertwo}
                                                  {\frac{3}{2}}{z}}{\GAMMA{\frac{3}{2}}\GAMMA{-\nuovertwo}}\right\}.
     \label{U_from_M}
  \end{equation}
It is important to note that Eq. (\ref{U_from_M}) is not a linear combination of two $M(a,b,z)$ functions.
Using Eq. (\ref{U_from_M}), it is possible to show that if $\nu$ is either zero or
 a positive integer number (denoted by $n$),  the solution in Eq. (\ref{sol_w}) can put  in the following form ($z>0$):
\footnote{This is easy to see from Eq. (\ref{U_from_M}) recalling that the Gamma function diverges at negative integer values.}
\begin{equation}
    w(z)\propto\left\{\begin{array}{ll}
                M(-\case{n}{2},\half,z)                 & \mathrm{for }~n~\mathrm{ even}\nonumber\\
                \sqrt{z}\,M(\half-\case{n}{2},\case{3}{2},z) & \mathrm{for }~n~\mathrm{ odd}\nonumber,\\
               \end{array}
         \right.
\end{equation}
or more compactly~\AS{b,c}, it can be written as $w(z)\propto H_n(\sqrt{z})$, with $H_n(z)$ the Hermite polynomial of order $n$; 
further more the product $w(z)e^{-|z|}$
converges for all $z$. Then, the full wave function has the usual form
$$
 \psi_n(y)\propto e^{-y^2/2}H_n(y),
$$
with the  corresponding eigenvalues being $\epsilon=n+\half$. What we have detailed  above condenses the
typical solution of the quantum harmonic oscillator  using special functions.

We now move to the solution of the quantum harmonic oscillator with a $\delta$-function potential  at the origin.
To that end, we have to review few properties of the $M(a,b,z)$ functions.
For non-integer values of $a$ and $b$, the $M(a,b,z)$ function is a convergent series for all finite given $z$~\AS{a},
but  diverges for $z\rightarrow+\infty$  as~\AS{d}
   \begin{equation}
       M(a,b,z)=\frac{\Gamma(b)}{\Gamma(a)}e^z z^{a-b}[1+O(|z|^{-1})].
    \label{lim_M_infty}
   \end{equation}
In terms of the original variable $y^2$, 
the function (\ref{lim_M_infty}) diverges as $e^{y^2}$, which implies that  $\psi(y)$ also
 diverges at infinity as $e^{y^2/2}$.
 Thus,  $\psi(y)$  is not,  in general, an acceptable wave function.

We noted at the introduction to this article that it is many times  referred (erroneously) 
in most standard textbooks on quantum mechanics
that the only mathematical solutions of the harmonic oscillator differential equation that does not blow up when $y\rightarrow+\infty$ are those
having  $\nu$ either zero or a positive integer.

On the contrary however, the function $U(-\nuovertwo,\half,y^2)$ with a 
non-integer $\nu$ does not blow up as $e^{y^2}$ when $y\rightarrow+\infty$. Indeed, it 
is easy to see [using Eqs.(\ref{U_from_M}) and (\ref{lim_M_infty})]
  that   $U(-\nu/2,1/2,y^2)\rightarrow y^\nu$~\AS{e} as $y\rightarrow+\infty$.
Thus, the function
\begin{equation}
\psi_\nu(y)=Ae^{-\half y^2}\,U(-\nuovertwo,\half,y^2)\,,
\label{eq_U_solution} 
\end{equation}
 could in principle be an acceptable wave function for any value of $\nu$, since it is both an even parity function of $y$
and square integrable (therefore normalizable). Why is it then that this solution 
has been casted away from textbooks? The weaker answer would be
because it does not provide quantized energy values, which is known to exist in any confined quantum system. 
The stronger answer is however that
the function $\psi_\nu(y)$ violates the boundary condition $\psi'_<(0)=\psi'_>(0)$ for any non-integer $\nu$, which 
must be obeyed by  the even parity wave functions $\psi(y)$  when $g=0$.  We are now about to see that
$\psi'_\nu(0^+)=-\psi'_\nu(0^-)$. It is the latter property of
$\psi_\nu(y)$ which allows the solution of the quantum problem (\ref{eq_hermite}) with finite $g$.

We have now gathered all the information needed to find the solutions of the eigenvalue problem (\ref{eq_hermite}).
Since the Hamiltonian (\ref{eq_Hamilt_HO}) is invariant over a parity  the transformation,
their eigenstates $\psi_g(y)$ are either even or odd parity states. In the case 
of odd states we have $\psi_g(0)=0$ and therefore they do not see the presence of the 
delta-function at the origin. Thus, the odd parity wave functions $\psi_g^{\rm odd}(y)$ are  the states
$\psi_n(y)$ of the ordinary harmonic oscillator, with $\nu=n=1,3,5,\ldots$ and the eigenvalues are
 $\epsilon=n+\half$. To latter result we call the ``trivial solution'' of the eigenproblem
(\ref{eq_hermite}).

\begin{figure}[ht]
\begin{center}
\includegraphics*[width=6cm]{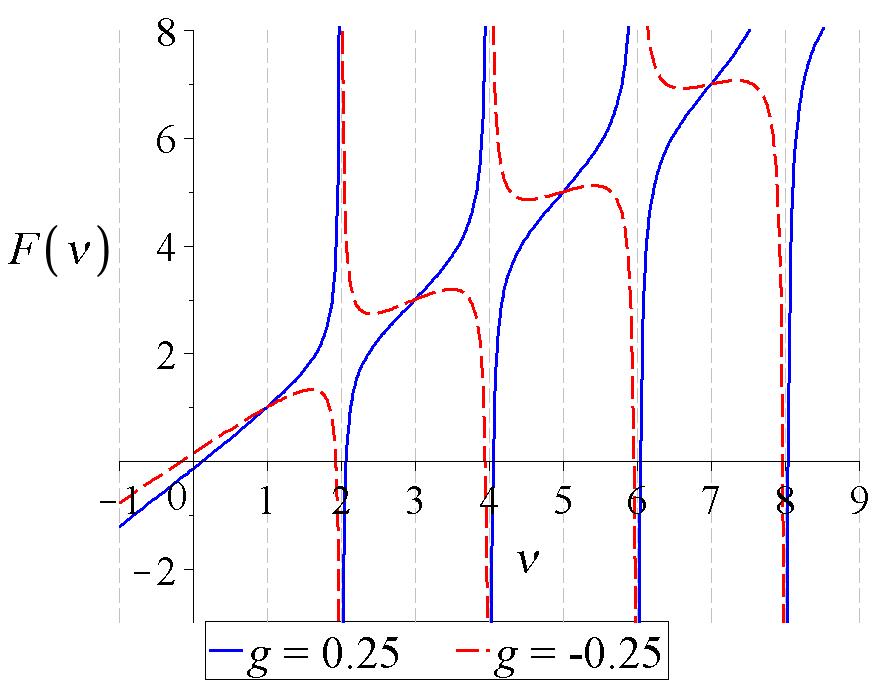}
\hspace{0.5cm}
\includegraphics*[width=6cm]{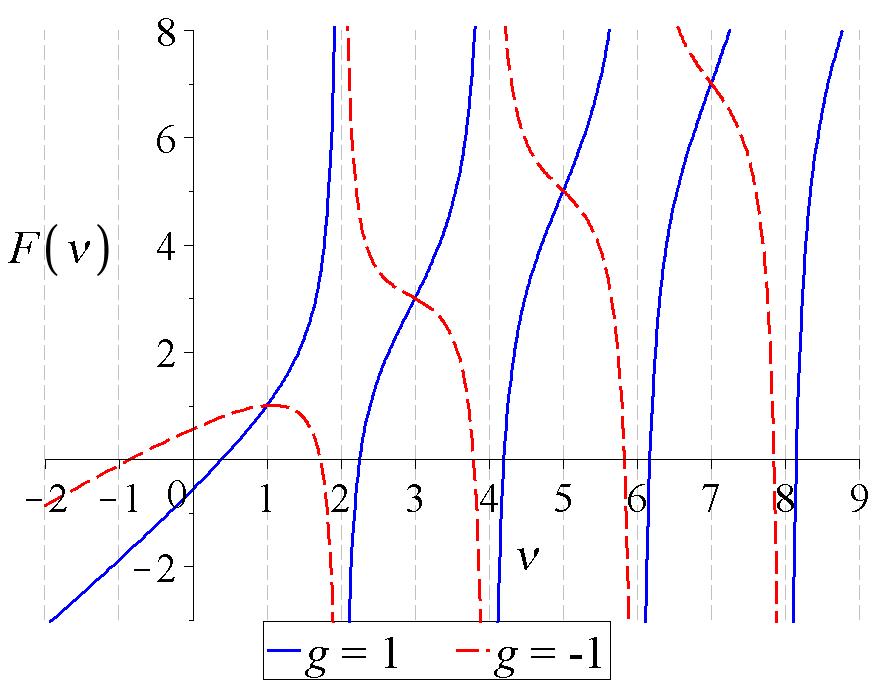}
\end{center}
\vspace{0.25cm}
\begin{center}
\includegraphics*[width=6cm]{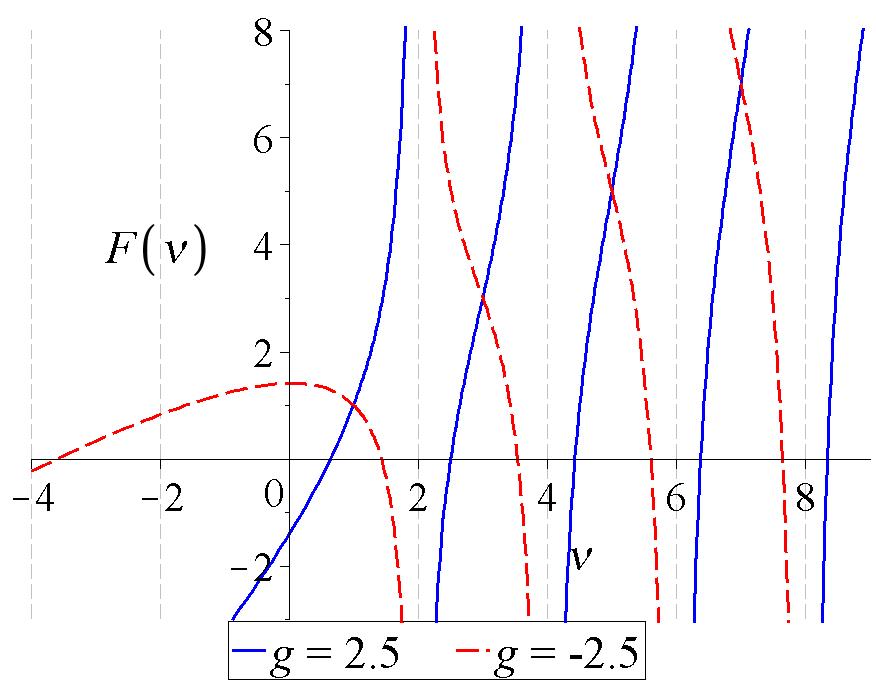}
\hspace{0.5cm}
\includegraphics*[width=6cm]{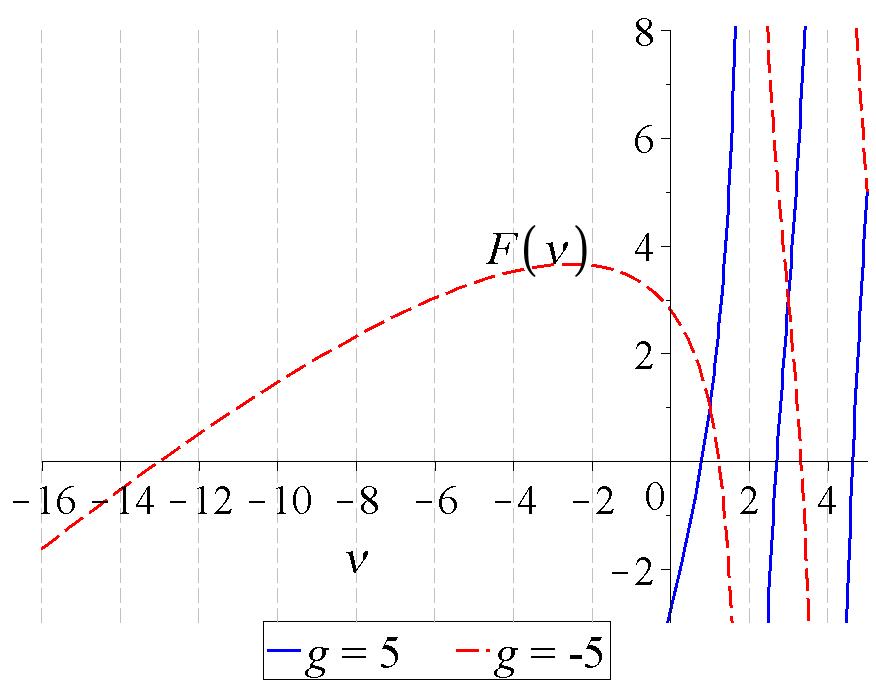}
\end{center}
\caption{(color online) Graphical solution of Eq. (\ref{eq_eigen_value_even}), 
for several positive and negative values of $g$. 
The numerical values of $\nu$ are given in Table \ref{tab_1}. The eigenvalues for
 the energy $E_\nu$ are given by $\nu+\half$ with $\nu$ the interceptions of the graphs
  with the $x$-axis (dashed lines refer to negative $g$, whereas solid lines refer to 
positive $g$).}
\label{fig_num_sol}
\end{figure}

\begin{figure}[ht]
\begin{center}
\includegraphics*[width=7cm]{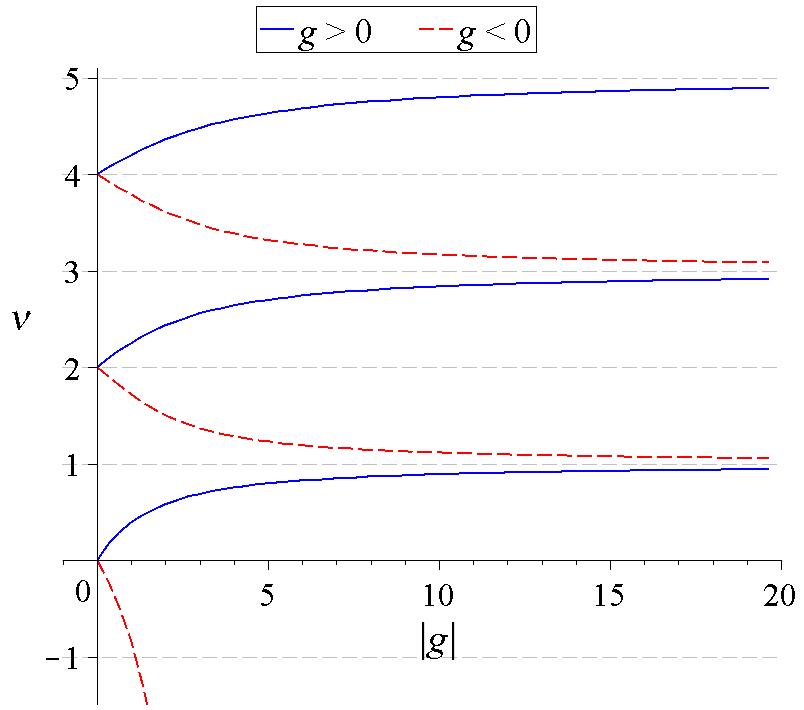}
\end{center}
\caption{(color online) Graphical solution of Eq. (\ref{eq_eigen_value_even}),
for several positive and negarive values of $g$. 
The numerical values of $\nu$ are given in Table \ref{tab_1}.}
\label{fig_nu_vs_g}
\end{figure}

The solution of the  even parity eigenfunctions $\psi_g^{\rm even}(y)$ is not as simple, since
these states feel the presence of the delta function at the origin. We need to find now the boundary condition the function
$w(y)$ must obey at the origin. Proceeding as we did at the introductory section, we integrate
Eq. (\ref{eq_hermite}) around $y=0$ obtaining:
\begin{equation}
  w'_>(0^+)-w'_<(0^-)=2gw(0)\,.
\label{eq_BC_HO_delta}
\end{equation}
Eq. (\ref{eq_BC_HO_delta}) enable us to find the quantized energies of the 
even parity eigenstates we are seeking.
Thus, the correct wave function for $\psi_g^{\rm even}(y)$ is $\psi_\nu(y)$
and not $\psi_n(y)$ with $n=0,2,4,\ldots$, as in the case $g=0$, for the
latter wave function violates the boundary condition (\ref{eq_BC_HO_delta}).
Using the results~\AS{f}
\footnote{Here, the second identity can be derived from the first one
 using the fact that~\AS{g} $U'(a,b,z)=-aU(a+1,b+1,z)$.}
\begin{eqnarray}
 \lim_{x\rightarrow0^+}U(-\nu/2,1/2,y^2)&=&\frac{\sqrt{\pi}}{\Gamma(1/2-\nu/2)}\,,\\
 \lim_{x\rightarrow0^+}U'(-\nu/2,1/2,y^2)&=&\frac{\nu\sqrt{\pi}}{\Gamma(1-\nu/2)}\,,
\end{eqnarray}
the eigenvalues associated with  even parity eigenstates of
Eq. (\ref{eq_Hamilt_HO_dless}) are given by the numerical solution of
the transcendent equation
\begin{equation}
F(\nu)\equiv\nu-g\frac{\Gamma(1-\nu/2)}{\Gamma(1/2-\nu/2)}=0\,,
\label{eq_eigen_value_even}
\end{equation}
which follows from the boundary condition (\ref{eq_BC_HO_delta}).
In Fig. (\ref{fig_num_sol}) we give the graphical solution of Eq. (\ref{eq_eigen_value_even}) for 
for $g=\pm0.25$, $g=\pm1.0$, $g=\pm2.5$, and $g=\pm5.0$, and in Table \ref{tab_1} the corresponding
numerical values of $\nu$, for the first five even-eigenstates.
 As expected, the effect of the potential is to shift the eigenenergies of the even-states of the ordinary harmonic oscillator
 up or down in energy  for positive and negative values of $g$, respectively. This effect is
 stronger for the low-lying eigenvalues (as we can anticipate from perturbation theory), and 
shifts  the eigenenergies of the states $\psi_\nu(x)$  toward those of their lower or higher neighboring
  odd-states, depending on the signal of $g$. This behavior is
 plotted in Figure \ref{fig_nu_vs_g}.
   In the problem we are dealing with, 
and contrary to the simple case of the ordinary harmonic oscillator, if $g<0$ there is also a negative  energy
eigenvalue, as we could have anticipated  from the solution of the attractive $\delta-$function potential
we have described in the introduction to this article. The absolute value of this negative energy state 
increases  with the strength of the $\delta-$function potential and, in the
  limit $g\rightarrow-\infty$, 
the confinement imposed by the harmonic potential becomes irrelevant and the wave function transforms into the bound state
  given by Eq. \ref{psiU_m0} and with the same eigenenergy. Indeed, using Stirling's formula~\AS{h}, it is easy to prove
  that as $z\rightarrow\infty$, $\Gamma(z+\half)/\Gamma(z)\rightarrow\sqrt{z}$. For $z=1+\half|\nu|$ this implies that
  as $g\rightarrow-\infty$, $g\sim\sqrt{2|\nu|}$. Since $E\sim -|\nu|\hbar\omega$
 \footnote{Note that in here, the harmonic oscillator frequency appears just as a by pass between the
            $\alpha$ and $g$ coupling constants: at the end, the harmonic potential, too swallow compared
           with the delta, plays no role in obtain result.},
  the former result implies that
  $$
    E_\nu=-g^2\half\hbar\omega=-\frac{\alpha^2 m}{2\hbar^2},
  $$
  which is the energy value we have  obtain before for the simple case of an isolated attractive $\delta-$function potential. 
\begin{table}[h]
\label{tab_1}
\begin{center}
 \begin{tabular}{c||r|r||r|r||r|r||r|r}
\hline
   $g\rightarrow0$&$-0.25$& $0.25$&$-1.0$& $1.0$ &$-2.5$ & $2.5$ & $-5.0$ & $5.0$ \\
\hline
             0.0 & -0.1557 &  0.1281 &-0.8424 &  0.3927 & -3.5865 &  0.6434 & -12.9900 &  0.7961 \\
             2.0 &  1.9288 &  2.0693 & 1.7208 &  2.2546 &  1.4285 &  2.5042 &   1.2305 &  2.7003 \\
             4.0 &  3.9469 &  4.0525 & 3.7912 &  4.2002 &  3.5420 &  4.4274 &   3.3227 &  4.6364 \\
             6.0 &  5.9558 &  6.0439 & 5.8258 &  6.1699 &  5.6051 &  6.3772 &   5.3833 &  6.5887 \\
             8.0 &  7.9614 &  8.0384 & 7.8473 &  8.1501 &  7.6473 &  8.3412 &   7.4285 &  8.5509 \\
\hline
\end{tabular}
\caption{Eigenvalues $\nu$ associated to the even parity eigenstates $\psi_\nu(y)$ for
several values of $g$; leftmost column corresponds to the case where $g\rightarrow0$, which coincide with the harmonic
 oscillator even parity eigenvalues; the ensuing columns stand for  $g=\pm0.25$,$g=\pm1.0$,$g=\pm2.5$ and $g=\pm5.0$.
In between each pair of eigenvalues we also have the states
associated with the odd parity eigenstates, which have eigenvalues
$\nu=1,3,5,7,\ldots.$ The associated eigenenergies are given by $\epsilon_\nu=\nu+\half$}
\end{center}
\end{table}

\begin{figure}[ht]
\begin{center}
\includegraphics*[width=7cm]{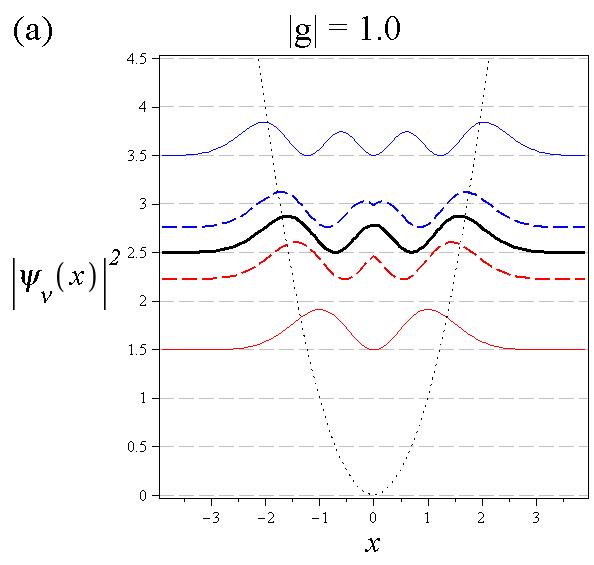}
\hspace{0.5cm}
\includegraphics*[width=7cm]{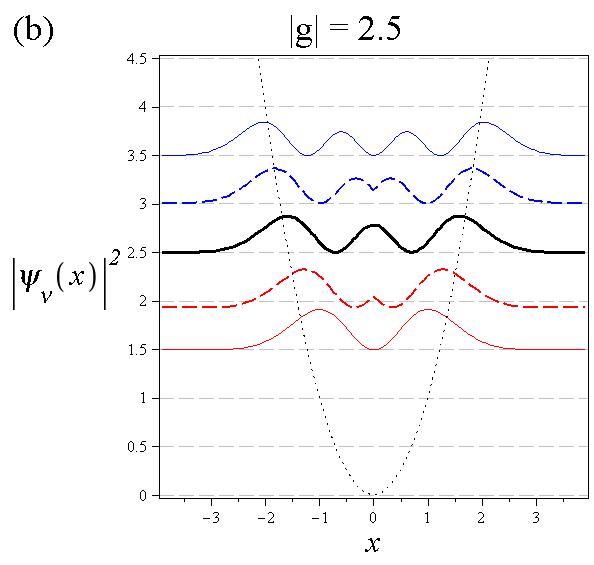}
\end{center}
\vspace{0.25cm}
\begin{center}
\includegraphics*[width=7cm]{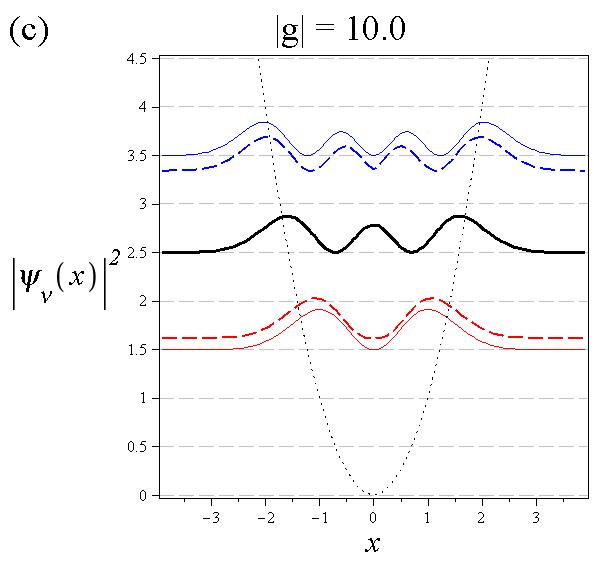}
\hspace{0.5cm}
\includegraphics*[width=7cm]{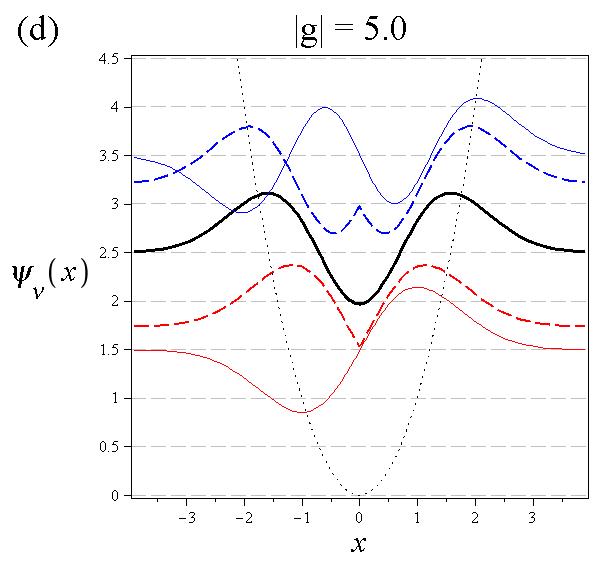}
\end{center}
\caption{(color online) Panels (a), (b) and (c): absolute square values of the wave functions
      of the harmonic oscillator (solid lines) for (from top to bottom) $\nu=n=3,2,1$
       (plotted, respectively, in blue, black and red)
      and for $\psi_\nu(x)$ (dashed lines) for $\nu$ corresponding to values
       of $g$ that varies from $1$ to $10$, for positive and negative values (top and bottom
       curves respectively). These plots show that the $|\psi_\nu(x)|^2$ curves approach the
       neighbors odd states. However, as it is shown in panel (d), the wave functions are quite different
        since the $\psi_\nu(x)$ are symmetrical with respect of $x$, presenting always a kink
         at $x=0$.}
\label{fig_wfs}
\end{figure}

Finally, in Figure \ref{fig_wfs} we plot in solid lines $\vert\psi_n(y)\vert^2$ as function of $y$
for $n=1,2,3$. In the panels (a), (b), and (c) of the same figure we also plot, in dashed lines,
 $\vert\psi_n(y)\vert^2$ for the wave functions that would correspond to the harmonic oscillator with $n=2$
  for different positive and negative values of $g$. As $g$ increases in positive (negative) values, the
  corresponding value of $\nu$ approaches $\nu=3$ ($\nu=2$) making  the
  absolute square  of the wave function, $|\psi_\nu(x)|^2$,   looking like the 
  absolute square of the wave function of its odd-parity state neighbor, $|\psi_{n\pm1}(x)|^2$.
  This doe not mean however that the two types of wave functions are the same, since they refer to orthogonal
eigenstates. To make this point evident, we plot both type of states in  panel (d) of
  Fig. \ref{fig_wfs}.

It is clear from Fig. \ref{fig_wfs} the behaviour $\psi'_\nu(0^+)\ne0$  (the number of nodes defines the order of the
state). When $g\gg1$, pairs of states (odd and even) of the harmonic oscillator with a delta-function
become quasi-degenerated. Indeed, in this regime the dip of the wave function $\psi_\nu(y)$ at $y=0$ approaches zero, 
but looking at Fig. \ref{fig_wfs} it is seen 
by the naked eye that the two functions are orthogonal (one is even and the other is odd; this is not self-evident
from the density probability graphs). The enhancement of the curvature of the wave function
around $y=0$ leads to an increase of the kinetic energy and therefore to an increase of energy
of the even-parity eigenstates.

\section{Conclusions}
We have discussed the solution of the Schr\"odinger equation for the one-dimensional 
harmonic potential with a Dirac delta-function at the origin. The odd-parity eigenstates
are given by the wave functions of the ordinary harmonic oscillator. This is obvious,
since the states of the latter system are zero at the origin and therefore do not feel 
the presence of the delta-function. For the even parity states the solution is non-trivial.
We have shown the existence of a solution of the differential equation of the harmonic 
oscillator that does not blow up at infinity as $e^{x^2/2}$ for non-integer values of $\nu$. 
As is well known, this solution is never
mentioned in quantum mechanics textbooks for a good reason,  but unfortunately that  reason  is,
as far as we know,
never discussed. Here we have shown that reason lies in the fact that its
derivative at $x=0$ is finite, violating the boundary conditions imposed in the simple harmonic
oscillator problem, that is, without the $\delta-$function at the origin. Nevertheless,
the wave function (\ref{eq_U_solution})
 is the one we need for solving the problem (\ref{eq_hermite}). In our work we have computed the 
eigenvalues and eigenfunctions of the even parity states 
and made, at the same time,  a little excursion to the  zoo of special functions
using only the famous {\it Handbook of Mathematical Functions} \cite{Abramowitze}.

{\it Note added:} After the submission of this article, the work
by  Busch {\it et al.} \cite{Busch1998}  was brought to our attention; 
the latter work elaborates 
on top of another paper by R. K. Janev and 
Z. Mari\'c \cite{Janev} and,
although focused on 
the three dimensional harmonic oscillator with a delta-function at the origin, one of the
figures in Ref. \cite{Busch1998} (Fig. 2) has the same information 
as our Fig. \ref{fig_nu_vs_g}, albeit presented in different form.
Further more, the energy eigenvalues of the zero angular momentum states
are given by an equation identical to our Eq. (\ref{eq_eigen_value_even}) but with
$1/2$ replaced by $3/2$ (for obvious reasons).


\section*{References}

\end{document}